\documentclass[prb,aps,twocolumn,showpacs,nofootinbib]{revtex4copy}
\usepackage{graphicx}% Include figure files
\usepackage{amsmath}
\usepackage{epstopdf}
\usepackage{amsbsy}
\usepackage{color}

\usepackage{bm} % bold math

\begin{document}

\title{Width of the longitudinal magnon in the vicinity of the O(3)
quantum critical point}
\author{Y. Kulik and O. P. Sushkov}
\affiliation{School of Physics, University of New South Wales, 
Sydney 2052, Australia}
\date{\today}

\begin{abstract}
We consider a three-dimensional quantum antiferromagnet in the vicinity of a 
quantum critical point separating the magnetically ordered and the magnetically
disordered phases. 
A specific example is TlCuCl$_3$ where the quantum phase
transition can be driven by hydrostatic pressure and/or by external 
magnetic field.
As expected two transverse and one longitudinal magnetic excitation have
been observed in the pressure driven magnetically ordered phase.
According to the experimental data, the longitudinal magnon has a substantial
width, which has not been understood and has remained a puzzle.
In the present work, we explain the mechanism for the width,
calculate the width and relate value of the width with parameters of
the Bose condensate of magnons observed in the same compound.
The method of an effective quantum field theory is employed
in the work.

\end{abstract}

\maketitle
\section{Introduction}
Continuous Quantum Phase Transitions (QPT) and behaviour of
quantum systems in vicinity of the corresponding quantum critical
points attract a great research interest\cite{Sachdevphystoday} both in theory and in 
experiment. There are manifestations
of such effects in cuprate superconductors, in iron pnictides and
in other materials. QPT are most pronounced in low dimensional systems,
but they also occur in three dimensional (3D) systems. In the present
paper we consider quantum systems, therefore 3D means 3D+time.
A well known material that manifests the 3D magnetic quantum critical behaviour
is TlCuCl${}_{3}$. Magnetic and other properties of the compound has been 
extensively  studied experimentally.~\cite{Takatsu97,Oosawa99,Nikuni,
Cav01,Choi03,Sherman03,
Glazkov04,Ruegg04extra,Ruegg,Tanaka04extra,Tanaka06Confextra,Tanaka07extra,
Tanaka08extra}
Under normal conditions the material is nonmagnetic, while a 
magnetic field and/or pressure drives a QPT to the magnetically ordered phase.
An analogous material KCuCl${}_{3}$ have similar 
properties.~\cite{Tanaka06Kcucl3Confextra,Tanaka07extraK}
It has been understood that TlCuCl${}_{3}$ consists of a 3D set of 
spin dimers
and the QPT in the magnetic field can be considered as Bose condensation
of the spin dimer triplet  excitations (triplons).~\cite{Mats02,Mats04,Misg04,
Sirker04,Sirker05,Sirker05a,Sachdevphystoday}
The pressure induced QPT can be also described in terms of
triplons.\cite{Norm04extra}
Inelastic neutron scattering from the magnetically ordered phase in the
vicinity of the pressure induced QPT~\cite{Ruegg} clearly indicates two transverse
and one longitudinal spin-wave excitation modes as one would expect for
condensed triplons.
According to the data~\cite{Ruegg} the longitudinal spin-wave excitation has a 
finite width. Generically this is also expected because of the possibility
for the longitudinal mode to decay to two transverse spin-wave excitations.
However to the best of our knowledge this width has never been calculated.
In the present work we perform such a calculation and compare the result
with the experimental data. The width is related to theparameters of
the Bose condensate of magnons observed in the same compound in the external 
magnetic field.
On the technical side in the present work we use the effective 
quantum field theory approach.

As we pointed out, a description of the system in terms of the spin dimer 
triplon technique is also possible. 
The spin dimer technique accounts for both the short range and the long 
range dynamics
on equal footing. On the one hand, this is a strength of this description. 
On the other hand, the description is quite technically involved 
if one needs to analyze the interaction effects that we are after.
The technical complexity is a price for the universality.
The effective quantum field theory method adopted in the present work,
allows us to account for the interaction effects much more accurately.
The method does not account for the short range dynamics, 
but the short range
effects are irrelevant in the vicinity of a QPT.

We would like also to note that in the present paper we use terms 
``spin waves'' and ``magnons'' on equal footing, having in mind magnetic
excitations on both sides of the QPT.

\section{Effective field theory}
The QPT that we consider here is a continuous zero temperature transition.
Therefore, the short range dynamics are not changed in the QPT, only the long
range dynamics are changed. In this situation it is quite natural 
to apply the effective field theory approach to describe the transition~\cite{sachdev10}. A similar approach was used to analyse the longitudinal mode in a quasi-one dimensional antiferromagnet.\cite{affleckWellman} Technically this 
approach is much simpler than the  triplon description.
The effective field theory must be written in terms of the vector field ${\vec \varphi}({t,\bf r})$ that describes staggered magnetization as well as
the long-wave-length magnetic excitations.
There is no net magnetization on both sides of the QPT, only the staggered magnetization, hence the  effective Landau-Ginzburg Lagrangian corresponding to
the nonlinear $\sigma$-model describes the system,
\begin{eqnarray} 
\label{LLs}
{\cal L}=\frac{1}{2}
\left({\dot{\vec \varphi}}-[{\vec \varphi}\times {\vec B}]\right)^2-
\frac{c^2}{2}\left({\bm \nabla}{\vec \varphi}\right)^2 -\frac{1}{2}m^2{\vec \varphi}^2
-\frac{1}{4}\alpha[{\vec \varphi}^2]^2\nonumber\\
\end{eqnarray}
We note the standard logic for this effective Lagrangian.
(i) The elastic energy must be proportional to $({\bm \nabla}{\vec \varphi})^2$ (stiffness).
(ii) The system consists of spins with the same gyromagnetic ratio $g=2$.
Therefore the Larmor theorem must be valid: 
If there is a state of the system without magnetic field,
then the same state in a uniform magnetic  field ${\vec B}$
must precess with 
frequency $\omega=B$ (Hereafter we set $g\mu_B=1$, where $\mu_B$ is Bohr magneton.)
(iii) The static energy can depend on the magnetic field only quadratically
because the system has no net magnetization. 
Note that instead of the hard constraint, ${\vec \varphi}^2=const$,
usually used in the nonlinear $\sigma$-model,  we impose the 
``soft'' quartic interaction $\frac{1}{4}\alpha[{\vec \varphi}^2]^2$.
The QPT is due to the mass term, 
\begin{equation}
\label{m}
m^2=\lambda^2(p_c-p) \ .
\end{equation}

In TlCuCl${}_{3}$ the QPT is driven by external hydrostatic pressure $p$ and this is reflected
in Eq.(\ref{m}), $\lambda^2 > 0$ is just a coefficient. In a more general situation pressure has to be
replaced by a generalized ``coupling constant''. 
When $p < p_c$ and there is no magnetic field, $B=0$, the mass squared is positive and this 
corresponds to
the magnetically disordered phase with the gapped triple degenerate spin-wave dispersion
$\omega_{\bm q}=\sqrt{c^2q^2+m^2}$. When $p > p_c$ the mass squared is negative and 
this results in a nonzero expectation value of $\langle{\vec \varphi}\rangle$ that
describes the spontaneous antiferromagnetic ordering with a gapped longitudinal mode
and two gapless Goldstone modes.

\section{Spin-wave dispersion in $\mbox{TlCuCl${}_{3}$}$ at zero pressure}
Since we will compare our results with experimental data for TlCuCl${}_{3}$ we need to determine
the corresponding spin-wave velocity $c$.  TlCuCl${}_{3}$ has a  monoclinic structure~\cite{Takatsu97}
with a unit cell very close to a rectangular prism with lattice parameters $a=3.97\AA$,
$b=14.13\AA$, $c=8.87\AA$ and an angle $\beta =96{}^\circ 10'$ only slightly deviating from a right 
angle ($\sin \beta =0.99421...$).
The spin wave dispersion over the Brillouin zone was determined by inelastic neutron scattering
and the dispersion can be very well fit by the following formula~\cite{Mats02,Sirker05a}
\begin{eqnarray}
\label{Js}
&& \epsilon_{\bm k}=\sqrt{(J+f_{\bm k}+g_{\bm k})^2-(f_{\bm k}+g_{\bm k})^2}\\
&&f_{\bm k}=J_a\cos k_x+J_{a2c}\cos(2k_x+k_z) \nonumber\\
&&g_{\bm k}=2J_{abc}\cos(k_x+k_z/2)\cos(k_y/2) \nonumber\\
&&J=5.52\text{ meV} \nonumber\\
&&J_a=-0.24\text{ meV} \nonumber\\
&&J_{a2c}=-1.57\text{ meV} \nonumber\\
&&J_{abc}=0.46\text{ meV}\ . \nonumber
\end{eqnarray}
Here $-\pi < k_x,k_y < \pi$, $-2\pi < k_z < 2\pi$ are dimensionless momenta. 
The dispersion has minimum at ${\bm k}_0=(0,0,2\pi)$. Near the minimum the dispersion squared can be expanded in powers of momentum, 
$\epsilon_{\bm k}^2=\Delta_0^2+a_1q_x^2+a_2q_y^2+a_3q_z^2+a_4q_xq_z+...$,
where $q_x=k_x$, $q_y=k_y$, $q_z=k_z-2\pi$, and 
 the coefficients $a_i$ are expressed in terms of parameters from (\ref{Js}).
Diagonalization of the quadratic form gives the spin-wave velocities
$c_1,c_2,c_3$,
\begin{eqnarray}
\label{c123}
&&\epsilon_{{\bm k}}^2=\Delta_0^2+c_1^2q_1^2+c_2^2q_2^2+c_3^2q_3^2\\
&&q_1=0.899q_x+0.438q_z\nonumber\\
&&q_2=q_y\nonumber\\
&&q_3=-0.438q_x+0.899q_z\nonumber\\
&&c_1=7.09\text{ meV}\nonumber\\
&&c_2=1.12\text{ meV}\nonumber\\
&&c_3=0.51\text{ meV}\nonumber\\
&&\Delta_0=0.65\text{ meV}\nonumber
\end{eqnarray}

The expansion (\ref{c123})  is generally valid only at small momenta, $q_1,q_2,q_3 \lesssim 1$, 
however,
accidentally the simple quadratic dependence along the ``softest'' direction $q_3$ 
is valid practically
up to the boundary of the Brillouine zone, $q_3\sim \pi$. Therefore, magnetism in TlCuCl${}_{3}$
is three-dimensional up to the energy/temperature $\epsilon \sim T \sim c_3\pi \sim 1.5\text{ meV}\approx 15\text{ K}$
above the gap. At a higher energy/temperature the magnetism becomes two-dimensional.

\section{Spin-wave excitations at zero temperature and zero magnetic field}
\begin{figure}[ht]
\includegraphics[width=0.4\textwidth,clip]{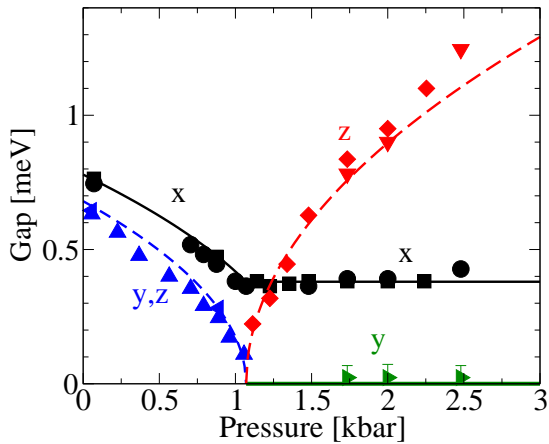}
\caption{({\it Color online}) Magnon gaps in the magnetically disordered
($p< p_c=1.07kbar$) and the magnetically ordered ($p> p_c$) phases.
The circles, diamonds, triangles etc, show experimental
data from Ref.\onlinecite{Ruegg}. The lines, solid and dashed show the theoretical
values of the gaps.
We take the convention that the staggered magnetization in the ordered phase is directed
along the z-axis, the corresponding longitudinal gap is shown by the topmost curve.
The Goldstone transverse y-magnon is gapless in the ordered phase, while the x-magnon is
slightly gapped due to the spin-orbit anisotropy.
}
\label{Figa}
\end{figure}

Taking into account Eq.(\ref{c123}) the stiffness term $\frac{c^2}{2}\left({\bm \nabla}{\vec \varphi}\right)^2$ in the effective action (\ref{LLs}) has to
be replaced by 
\begin{eqnarray}
\label{cr}
\frac{1}{2}\left[c_1^2(\partial_1{\vec \varphi})^2+c_2^2(\partial_2{\vec \varphi})^2
+c_3^2(\partial_3{\vec \varphi})^2\right]\ .
\end{eqnarray}
In the magnetically disordered phase, $p < p_c$ the Lagrangian (\ref{LLs}) yields the following
triple degenerate spectrum
\begin{eqnarray}
\label{sp1}
\omega_{\bm q}&=&\sqrt{\Delta^2+c_1^2q_1^2+c_2^2q_2^2+c_3^2q_3^2}\\
\Delta&=&\sqrt{m^2}=\lambda\sqrt{p_c-p}\ . \nonumber
\end{eqnarray}
Comparing with the gap measured in neutron scattering, Ref.~\cite{Ruegg}, and presented in Fig.~\ref{Figa}
by blue triangles we determine the value of $\lambda$, $\lambda = 0.66\text{ meV/kbar}^{1/2}$.
Here we take into account that $p_c=1.07\text{ kbar}$.
%%%%%%%%%%%%%%%%%
%%%%%%%%%%%%%%%%%
The experimental data~\cite{Ruegg} presented in Fig.~\ref{Figa} demonstrate some anisotropy 
that is not accounted
in the Lagrangian (\ref{LLs}). At the moment we disregard the anisotropy and return to this 
point later.
On the other side of the QPT, $p > p_c$, the field ${\vec \varphi}$ attains spontaneously
the following expectation value
\begin{equation}
\label{vf}
|\varphi_c|=\sqrt{\frac{|m^2|}{\alpha}}=\frac{\lambda}{\sqrt{\alpha}}\sqrt{p-p_c} \ .
\end{equation}
The subscript ``c'' in $\varphi_c$ stands for classical expectation value.
We chose the classical field to be directed along the z-axis in the spin space. 
Further standard analysis of the effective Lagrangian shows that there are two
transverse gapless Goldstone x- and y-modes 
\begin{eqnarray}
\label{spt}
\omega_{\perp\bm q}=\sqrt{c_1^2q_1^2+c_2^2q_2^2+c_3^2q_3^2}\ ,
\end{eqnarray}
and one gapped longitudinal z-mode with dispersion
\begin{eqnarray}
\label{sp2}
\omega_{L\bm q}&=&\sqrt{\Delta^2_L+c_1^2q_1^2+c_2^2q_2^2+c_3^2q_3^2}\\
\Delta_L&=&\sqrt{2|m^2|}=\sqrt{2}\lambda\sqrt{p-p_c}\ . \nonumber
\end{eqnarray}
The ``longitudinal'' gap agrees pretty well with the data shown in Fig.~\ref{Figa} by red 
triangles and diamonds.

\section{Bose condensation in magnetic field and the quartic term in the Lagrangian}
To calculate the width of the longitudinal spin wave we need to know the coefficient
$\alpha$ in the quartic term in (\ref{LLs}). To determine the coefficient we turn to the
analysis of Bose condensation of magnons in the external magnetic field at zero pressure.
The Lagrangian (\ref{LLs}) results in the following energy density
\begin{eqnarray}
\label{en}
 E & = & 
{\dot {\vec \varphi}} \frac{\delta L}{\delta {\dot {\vec \varphi}}}-L\nonumber\\
&\to&\frac{1}{2} \left(m^{2} -B^{2} \right)\varphi^{2} +\frac{1}{2} ({\vec B}\cdot{\vec \varphi})^2 
+\frac{1}{4} \alpha [{\vec \varphi} ^{2}]^2\ .
\end{eqnarray}
Here $m^2> 0$ since $p < p_c$.  
In the energy we skip all terms with derivatives since we are interested in Bose condensation in
a homogeneous system.
For magnetic field $B<m$, the energy is minimized when $\vec{\varphi }=0$, there is no spontaneous magnetization.
For magnetic field $B>m$ (i.e. when the gap is closed), the energy is minimum when 
\begin{eqnarray}
\label{12}
&&({\vec \varphi}\cdot{\vec B})=0 \nonumber\\
&&\varphi_{\perp}^{2} =(B^{2} -m^{2})/\alpha \ .
\end{eqnarray}
Thus, there is a spontaneous 
staggered magnetization perpendicular to the direction of the magnetic field.
The magnetization can be arbitrarily rotated around the direction of the magnetic field, therefore,
the system retains a U(1) symmetry as it should be for a Bose condensate.
Substitution of the expectation value (\ref{12}) back to Eq.(\ref{en}) gives the
following energy density of the condensate
\begin{eqnarray}
\label{en3}
E = -\frac{1}{4} \frac{\left(B^{2} -m^{2} \right)^{2} }{\alpha }\approx
-\frac{m^{2} }{\alpha } \left(B-m\right)^{2}
\end{eqnarray}
In the second equality in this equation we assume that $B-m \ll m$, i.e. we are close to the phase transition point at $B=m$.
This concludes the analysis of Bose condensation in terms of the effective action.

It is well known that the Bose condensation of magnons can be also analysed 
in terms of usual Schrodinger equation within the Hartree-Fock-Popov
approximation.
Such an analysis has been performed earlier in Ref. \onlinecite{Sirker05a}.
In the Hartree-Fock-Popov approximation the density energy of the 
condensate at zero temperature reads
\begin{equation}
\label{HFP}
E=\left(m-B\right)n_{0} +\frac{v_{0} }{2} n_{0}^{2}\to
-\frac{1}{2} \frac{\left(B-m\right)^{2} }{v_{0} }
\end{equation}
where $n_0$ is the condensate density and $v_0$ is the
 effective repulsion between magnons. The arrow ``$\to$'' shows the
result of the energy minimization with respect to $n_0$.
Comparing (\ref{en3}) and (\ref{HFP}) we express the field theory
parameter $\alpha$ is terms of the Hartree-Fock-Popov parameter $v_0$,
$\alpha =2v_{0} m^{2}$.
Fit of data~\cite{Nikuni}  for Bose condensation in TlCuCl$_3$ 
performed  in Ref.~\onlinecite{Sirker05a} gave the value of the
effective repulsion, $v_0\approx 25\text{ meV}$. The data in Ref. \onlinecite{Nikuni} is taken
at zero pressure, therefore $m=\Delta_0\approx 0.65\text{ meV}$. All in all we get
\begin{equation}
\label{av}
\alpha =2v_{0} m^{2}\approx 21\text{ meV}^3\ .
\end{equation}
According to the fit in Ref. \onlinecite{Sirker05a}, there is a margin of error of about 20\% in 
the values of $v_0$ and $\alpha$.

\section{Width of the longitudinal magnon}
%%%%%%%%%%%%%%%%%
\begin{figure}[ht!]
\includegraphics[width=0.2\textwidth,clip]{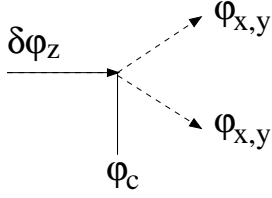}
\caption{Decay amplitude of the longitudinal magnon.
The dashed lines indicate magnons and the solid line
indicates the condensate field $\varphi_{c}$.
$\delta\varphi_z$ corresponds to the longitudinal magnon,
$\varphi_{\perp}$ corresponds to transverse Goldstone magnons.
}
\label{FigLd}
\end{figure}
%%%%%%%%%%%%%%%%%
Decay of the longitudinal magnon is due to the quartic term
in (\ref{LLs}) and is described by the diagram shown in Fig.~\ref{FigLd}.
To calculate the decay amplitude one needs first to rewrite the Lagrangian 
in terms of magnons in the magnetically ordered phase.
After substitution of  $\varphi_z=\varphi_c+\delta\varphi_z$ in Eq.(\ref{LLs})
(the condensate field $\varphi_c$ is given by Eq.(\ref{vf})) we find the
quadratic and the cubic part of the Lagrangian
\begin{eqnarray}
\label{L23}
L&=&L_2+L_3+L_4\\
L_2&=&
\frac{1}{2}
\left({\dot{\vec \varphi}}_{\perp}\right)^2-
\frac{c^2}{2}\left({\bm \nabla}{\vec \varphi}_{\perp}\right)^2\nonumber\\
&+&\frac{1}{2}
\left({\dot{\delta \varphi}}_z\right)^2-
\frac{c^2}{2}\left({\bm \nabla}{\delta \varphi}_{z}\right)^2 -\frac{1}{2}|2m^2|({\delta \varphi}_z)^2
\nonumber\\
L_3&=&-\alpha\varphi_c\ \delta\varphi_z({\vec \varphi}_{\perp})^2\nonumber
\end{eqnarray}
The stiffness term $\frac{c^2}{2}\left({\bm \nabla}...\right)^2$ is determined in Eq.(\ref{cr}).
Standard quantization of the quadratic Lagrangian gives (we set $\hbar=1$)
\begin{eqnarray} 
\label{rw}
\delta\varphi_z&=&\sum_{\bf k}\frac{1}{\sqrt{2\omega_{L\bf k}}}\left(
e^{i\omega_{L\bf k}t-i{\bf k}\cdot{\bf r}}a_{L\bf k}^{\dag}+
e^{-i\omega_{L\bf k}t+i{\bf k}\cdot{\bf r}}a_{L\bf k}\right)\nonumber \\
\varphi_{x,y}&=&\sum_{\bf q}\frac{1}{\sqrt{2\omega_{\perp\bf q}}}\left(
e^{i\omega_{\perp\bf q}t-i{\bf q}\cdot{\bf r}}a_{x,y,\bf q}^{\dag}+
e^{-i\omega_{\perp\bf q}t+i{\bf q}\cdot{\bf r}}a_{x,y,\bf q}\right) \ , \nonumber
\end{eqnarray}
where $a$ and $a^{\dag}$ are corresponding annihilation and creation operators.
The decay matrix element shown in Fig.~\ref{FigLd} is determined after the operators
(\ref{rw}) are substituted in the cubic Lagrangian $L_3$,
\begin{eqnarray}
\label{M}
M=\frac{2\alpha\varphi_c}{\sqrt{2\omega_{L\bf q}2\omega_{\perp\bf q_1}2\omega_{\perp\bf q_2}}} \ ,
\end{eqnarray}
where ${\bf q}$ is the momentum of the initial longitudinal magnon and 
${\bf q}_1$, ${\bf q}_2$ are the momenta of the final transverse magnons.
The coefficient 2 in (\ref{M}) is due to two different ways for pairing of
creation and annihilation operators.
The decay width is given by Fermi golden rule
\begin{eqnarray}
\label{gr}
\Gamma&=&\frac{1}{2}(2\pi)^4\sum_{x,y}\int\frac{d^3q_1}{(2\pi)^3}\frac{d^3q_2}{(2\pi)^3}
\nonumber\\
&\times&
|M|^2\delta(\omega_{L\bf q}-\omega_{\perp\bf q_1}-\omega_{\perp\bf q_2})\delta({\bm q}-{\bm q}_1-{\bm q}_2)
\end{eqnarray}
where the coefficient $1/2$ stands to avoid the double counting of final bosonic states.
Integration in (\ref{gr}) is straightforward, the answer is
\begin{eqnarray}
\label{gr1}
&&\Gamma_q=\frac{\alpha}{8\pi c_1c_2c_3}
\frac{\Delta_L^2}{\sqrt{\Delta^2_L+c_1^2q_1^2+c_2^2q_2^2+c_3^2q_3^2}}\nonumber\\
&&\Gamma_{q=0}=\frac{\alpha}{8\pi c_1c_2c_3}\Delta_L\ . 
\end{eqnarray}
In this equation $q_1,q_2,q_3$ are the components of the vector ${\bm q}$.
%%%%%%%%%%%%%%%%%
\begin{figure}[ht]
\vspace{10pt}
\includegraphics[width=0.4\textwidth,clip]{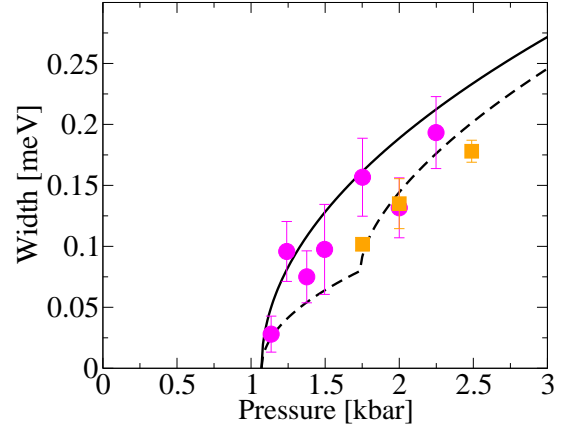}
\caption{
({\it Color online}) The decay width of the longitudinal magnon at $q=0$.
The circles and squares, show experimental
data from Ref.\onlinecite{Ruegg}. The solid line shows the theoretical
value calculated without account of the spin-orbit anisotropy.
The dashed line shows the theoretical
value with account of the spin-orbit anisotropy.
}
\label{Figb}
\end{figure}
%%%%%%%%%%%%%%%%%
According to (\ref{gr1}) the width $\Gamma_{q=0}\propto \Delta_L\propto \sqrt{p-p_c}$
in a reasonable agreement with experimental data from Ref.\onlinecite{Ruegg} presented in 
Fig.\ref{Figb}.
According to Eqs. (\ref{gr1}),(\ref{c123}),(\ref{av}) the theoretical value
for the ratio $\Gamma_{q=0}/\Delta_L$ is $\Gamma_{q=0}/\Delta_L=
\frac{\alpha}{8\pi c_1c_2c_3}\approx 0.21$.
The corresponding theoretical curve for $\Gamma_{q=0}$ versus pressure is shown 
in Fig.~\ref{Figb} by the black solid line together with experimental data~\cite{Ruegg} shown in the same figure.
We stress that we calculate not only the functional dependence of the gap on the pressure.
We calculate the absolute value of the gap. 
Overall the agreement between the theory and the experiment is quite good
having in mind experimental error bars, Fig.\ref{Figb}, as well as 20\% theoretical
uncertainty in the value of $\alpha$.
The agreement can be further improved if one accounts for the spin-orbit anisotropy.
We discuss the anisotropy in the next section.

\section{Account for anisotropy induced by the spin-orbit interaction}
Data presented in Fig.\ref{Figa} clearly indicate an easy plane anisotropy in
the magnetically ordered phase: one of the transverse magnons is gapped with the
constant gap
\begin{equation}
\label{dso}
\Delta_{a} = 0.38 \text{ meV}\,
\end{equation}
where the subscript ``a'' stands for anisotropy.
To incorporate the anisotropy in our description we have to add the term
\begin{equation}
\label{la}
\delta L_a=\frac{1}{2}\Delta_a^2\varphi_x^2
\end{equation}
to the effective Lagrangian (\ref{LLs}).
With account of the anisotropy the triple degeneracy of the dispersion
in the magnetically disordered phase, $p < p_c$, is lifted, the z- and y-modes
have the same dispersion (\ref{sp1}) as before while the x-mode has a higher energy
\begin{eqnarray}
\label{sp1a}
x: && \omega_{\bm q}=\sqrt{\Delta^2+\Delta_a^2+c_1^2q_1^2+c_2^2q_2^2+c_3^2q_3^2}\nonumber\\
y,z: && \omega_{\bm q}=\sqrt{\Delta^2+c_1^2q_1^2+c_2^2q_2^2+c_3^2q_3^2}\nonumber\\
&&\   \Delta=\sqrt{m^2}=\lambda\sqrt{p_c-p}\ . \nonumber
\end{eqnarray}
The corresponding gaps ($\omega_{q=0}$) are plotted by lines in Fig.\ref{Figa} versus pressure.
The dispersion in the magnetically ordered phase, $p> p_c$,  is the following
\begin{eqnarray}
\label{sp2a}
x: && \omega_{\bm q}=\sqrt{\Delta_a^2+c_1^2q_1^2+c_2^2q_2^2+c_3^2q_3^2}\nonumber\\
y: && \omega_{\bm q}=\sqrt{c_1^2q_1^2+c_2^2q_2^2+c_3^2q_3^2}\nonumber\\
z: && \omega_{\bm q}=\sqrt{\Delta_L^2+c_1^2q_1^2+c_2^2q_2^2+c_3^2q_3^2}\nonumber\\
&& \   \Delta_L=\sqrt{2|m^2|}=\sqrt{2}\lambda\sqrt{p_c-p}\ .
\end{eqnarray}
The corresponding gaps ($\omega_{q=0}$) are also plotted by lines in Fig.\ref{Figa} versus pressure.
Overall agreement between the theory and the experimental data presented in the same
Fig.\ref{Figa} is excellent.

The anisotropy gap influences the width of the longitudinal magnon only via limitation
of the decay phase space.  
A straightforward calculation shows that Eq.(\ref{gr1}) has to be
modified in the following way
\begin{eqnarray}
\label{gr1a}
\Gamma_{q=0}=\frac{\alpha}{16\pi c_1c_2c_3}\Delta_L\left[1+\frac{\sqrt{\Delta_L^2-4\Delta_a^2}}{\Delta_L}
\theta\left(\Delta_L^2-4\Delta_a^2\right)\right] \ .
\end{eqnarray}
%%%%%%%%%%%%%%%%%
  Here $\theta(x)$ is the step function, $\theta(x)=1$ if $x > 0$, otherwise $\theta(x)=0$.
The calculated decay width is shown in Fig.\ref{Figb} by the dashed line.
The agreement with experiment is excellent.

\section{conclusions}
The decay width of the longitudinal magnon in the vicinity of the O(3)
quantum critical point of a three-dimensional quantum antiferromagnet has been 
calculated using the effective field theory approach.
The results are quite generic.

We also compared our results with experimental data for
TlCuCl$_3$ where the quantum phase
transition can be driven by hydrostatic pressure and/or by external 
magnetic field.
The parameters of the effective action were determined from available data
for inelastic neutron scattering
and for Bose condensation of magnons in the external magnetic field.
Having determined these parameters, we calculated the decay width.
The result agrees very well with measurements.

\section{acknowledgement}
We are grateful to B. Normand for stimulating discussions
and to J. Oitmaa for important comments.

\end{document}